\documentclass[%
 reprint,
superscriptaddress,
 amsmath,amssymb,
 aps,
 prl,
]{revtex4-2}

\usepackage{graphicx}
\usepackage{dcolumn}
\usepackage{bm}
\usepackage[colorlinks= true, linkcolor=black, citecolor=black, urlcolor=black]{hyperref}
\usepackage{lipsum}

\usepackage{xcolor}

\begin{document}


\title{Asymptotic absorption-time distributions in extinction-prone Markov processes}

\author{David Hathcock}
\affiliation{Department of Physics,  Cornell University, Ithaca, New York 14853, USA}
\author{Steven H. Strogatz}
\affiliation{Department of Mathematics,  Cornell University, Ithaca, New York 14853, USA}

\date{\today}

\begin{abstract}

We characterize absorption-time distributions for birth-death Markov chains with an absorbing boundary. {\color{black}For ``extinction-prone'' chains (which drift on average toward the absorbing state) the asymptotic distribution is Gaussian, Gumbel, or belongs to a family of skewed distributions. The latter two cases arise when the dynamics slow down dramatically near the boundary.} Several models of evolution, epidemics, and chemical reactions fall into these classes; in each case we establish new results for the absorption-time distribution. Applications to African sleeping sickness are discussed.
\end{abstract}

\pacs{Valid PACS appear here}
\maketitle

 Modeling extinction-prone dynamics is essential to our understanding of epidemics, disease incubation, and evolution. For example, a key goal in epidemiology is to implement control measures (such as social distancing or vaccination) that push the dynamics toward a state where the disease is eradicated on a reasonable timescale \cite{hinman1998global, hopkins2013disease, aliee2020estimating}. Similarly, disease incubation \cite{sartwell1950distribution, ottino2017evolutionary} and evolution \cite{nowak2006evolutionary, lieberman2005evolutionary} involve highly fit infectious cells or mutant species outcompeting their less fit counterparts.

In these fields the distribution of extinction times, rather than just the mean, is crucial. For example, how long must a patient wait after exposure to a disease to be sure they are not infected? In the best and worst case scenarios, how long must epidemiological control measures be imposed to stop an outbreak? Knowledge of the extinction-time distribution provides an answer to these questions. Incubation period distributions have long been measured empirically to inform treatment regimens or public health initiatives \cite{sartwell1950distribution}.  Similarly, a recent study used a data-driven model of African sleeping sickness in the Democratic Republic of Congo to predict the distribution of times until the disease is eradicated \cite{aliee2020estimating}.

In this Letter, we show that two particular extinction-time distributions---Gaussian and Gumbel distributions---arise generically from basic features of the stochastic dynamics driving the system. 
These distributions were found previously in several models of evolutionary dynamics \cite{ottino2017evolutionary, ottino2017takeover, hathcock2019fitness}. 
We show now that these same distributions appear in much more general classes of birth-death Markov chains, {\color{black} along with a family of skewed distributions that include the Gumbel}. Extending the approach introduced in Ref.~\cite{hathcock2019fitness}, we provide analytical criteria that predict when the asymptotic absorption-time distribution is normal, Gumbel,{ \color{black} or a member of the family of skewed distributions.}
We apply our results to models of epidemiology \cite{pastor2015epidemic, dorogovtsev2008critical, jacquez1993stochastic}, ecology \cite{grasman1997local, azaele2016statistical, gandhi1998critical}, stochastic chemical reactions \cite{doering2007asymptotics, van1992stochastic}, and evolutionary games \cite{ meyer2020evolutionary}, for which the predicted distributions agree with those measured via simulation. To our knowledge, this is the first calculation of the asymptotic absorption-time distributions for these models.  As an application, we show that the Gumbel distribution closely resembles eradication-time distributions for African sleeping sickness. 

We analyze birth-death Markov processes with a linear chain of states $m=0, 1, \dots, N$. For example, $m$ might represent the number of infected individuals in an epidemic. The system has an absorbing state at $m=0$ (where nobody is infected) and a reflecting state at $m=N$ (the maximum allowed infected population). Transitions occur only between neighboring states, i.e., the population can only increment by 1 in either direction. The dynamics of $p_m(t)$, the probability of occupying state $m$ at time $t$, obey the master equation,
\begin{equation}
\dot p_m(t)  = b_{m-1} p_{m-1}(t) + d_{m+1} p_{m+1}(t) - (b_m + d_m) p_m(t),
\end{equation}
where $b_m$ and $d_m$ are respectively the birth and death rates at which the state increases or decreases from state $m$. The master equation can also be expressed as $\dot{\mathbf{p}}(t) = \Omega \cdot \mathbf{p}(t)$, where $\Omega$ is the transition matrix containing the birth and death rates. Since the state at $m=0$ is absorbing and the state $m=N$ is reflecting, we have $b_0=b_N=0$. For simplicity we assume the system starts in an initial state $m=N$, i.e. $p_m(0) = \delta_{m,N}$, but our results apply more broadly \cite{SM}. The quantity we are interested in is the first-passage time $T$ to the absorbing state $m=0$; here we focus on obtaining the probability distribution about the mean. 


Building on our recent results \cite{hathcock2019fitness}, we develop an approach to determine the absorption-time distributions for general classes of birth-death Markov chains in the limit of large system size. The key insight is to introduce a change of variables, $D_m = b_m + d_m$ and $r_m = b_m/d_m$. If the system is in state $m$, it waits on average a time $D_m^{-1}$ before increasing or decreasing. The probabilities of the next step being forward or backward are $r_m/(1+r_m)$ and $1/(1+r_m)$ respectively; $r_m$ is the ratio of these probabilities. Thus, our coordinate change separates the random-walk portion of the Markov process, which describes the relative probabilities of stepping forward or backward at each state, from the times spent waiting in each state. This change of variables leads to a transition matrix decomposition, $\Omega =  \Omega_{RW} D$, where $D$ is diagonal with elements $D_m$  and $\Omega_{RW}$ is the transition matrix for a biased random walk. The number of times the system visits each state depends only on the random-walk portion of the process. The elements $V_{ij}$ of $V=-\Omega_{RW}^{-1}$ encode the average number of visits to state $i$ before absorption, starting from state $j$.

To characterize the asymptotic distributions, we compute the cumulants $\kappa_n(N)$ of the absorption time $T$, which describe the shape of the distribution. For instance, $\kappa_1$ is the mean, $\kappa_2$ is the variance, and $\kappa_3/\kappa_2^{3/2}$ is the skew. Following Ref.~\cite{hathcock2019fitness} we use the matrix decomposition above to derive the  cumulants (generalizing the previous result to non-constant $r_j$):
\begin{equation}\label{exactCumulants}
\color{black}    \kappa_n(\{r_j\}, N) = \sum_{1\leq i_1 \leq i_2 \leq \dots \leq i_n \leq N} \frac{w_{i_1i_2 \cdots i_n}^n(\{r_j\})}{(b_{i_1}+d_{i_1})\cdots(b_{i_n}+d_{i_n})}.
\end{equation}
Here $w_{i_1i_2 \cdots i_n}^n(\{r_j\})$ are weighting factors that depend only on the visit statistics of the random walk; for example, $w_i^1(\{r_j\}) = V_{i, i}$. See the Supplemental Material \cite{SM} for a derivation of this formula and explicit expressions for the first few weighting factors, each of which are polynomials of the visit numbers $V_{ij}$. {\color{black} Equation~(\ref{exactCumulants}) is equivalent to well known recursive relations for absorption time moments \cite{goel2016stochastic}, but this form enables the asymptotic analysis leading to the results below.}

{\color{black} The weighting factors have some convenient properties.}
First, they appear to be non-negative: $w_{i_1i_2 \cdots i_n}^n(\{r_j\}) \geq 0$ and \emph{increasing} functions of each $r_j$. We show the non-negativity and monotonicity explicitly up to order $n=4$ \cite{SM} and conjecture these properties hold for all orders. 
Second, the weighting factors appear to fall off exponentially away from the diagonal. For constant $r_j=r$, this exponential decay can be shown explicitly \cite{hathcock2019fitness}. We conjecture that the same decay holds for arbitrary transition probabilities $\{r_j\}$. The intuition is that the visits to state $i$ are uncorrelated with those to state $j$ (for $N\gg1$ and $i-j = \mathcal{O}(N)$), due to the Markov property. 
%

The first universality class of birth-death Markov chains we consider have normally distributed absorption times. {\color{black} As an instructive special case, consider the process $b_m = 0$, $d_m = d$, which visits each state exactly once before absorption, waiting a time $d^{-1}$ on average at each. The time to absorption is simply $T = \sum_m \mathcal{E}_m(d)$ where $\mathcal{E}_m(d)$ is an exponential random variable. Since $T$ is a sum of identical random variables we expect it to be normally distributed by the central limit theorem. Alternatively, the cumulants of $T$ are $\kappa_n = N/d^n$. In units of the standard deviation the higher order  cumulants vanish: $\kappa_n/\kappa_2^{n/2} = N^{1-n/2} \to 0$ as $N \rightarrow \infty$. Hence the distribution is asymptotically normal. 

We might also expect this asymptotic normality to hold for transition rates with mild state dependence: if $b_m+d_m$ does not vary too much (we will give a precise condition below), the absorption time is a sum of nearly identical exponential random times. Similarly, for $r_m=b_m/d_m>0$, the system randomly walks back and forth, but as long as $r_m<1$ the average number of visits to each state is finite. Under either of these generalizations the distribution is asymptotically normal. 
}

To characterize more precisely which Markov chains lead to normally distributed absorption times, we compute the asymptotic form of the cumulants in Eq.~(\ref{exactCumulants}) by introducing two auxiliary Markov chains. These have the same $b_i+d_i$ as the original system, but $b_i$ and $d_i$ are adjusted so that the ratios are  $r_j = r_{\max}$ or $r_j = r_{\min}$, where $r_{\max} = \lim_{N\rightarrow\infty} \max_{1<j<N} r_j$ and $r_{\min} = \lim_{N\rightarrow\infty} \min_{1<j<N} r_j$. In other words, we construct two Markov chains where the time spent waiting in each state is identical to that for the original system, but the odds of moving toward the absorbing state are increased or decreased to be uniform. 

Above we noted that the weighting factors $w^n$ in  Eq.~(\ref{exactCumulants}) are increasing functions of $r_j$. Thus, we can bound the cumulants in our system by those for the auxiliary Markov chains, $\kappa_n(r_{\min}, N) \leq \kappa_n(\{r_j\}, N) \leq \kappa_n(r_{\max}, N)$. The asymptotic form of $\kappa_n(r,N)$ (where $r$ is constant across states) was computed in Ref.~\cite{hathcock2019fitness}; we summarize the calculation in the Supplemental Material \cite{SM}. To nail down the asymptotics of $\kappa_n(r,N)$ we require the waiting times to be `flat' in the following sense:
\begin{equation}\label{universalityCondition}
\color{black} \frac{1}{N}\sum_{m=1}^N t_m \sim c \max_{1\leq m \leq N} t_m,
\end{equation}
 {\color{black}  where $t_m= (b_m+d_m)^{-1}$ is the mean waiting time at state $m$ and  $c$ is a constant independent of $N$. In other words, the mean waiting time $\langle t_m\rangle$ across all states is the same asymptotic order as the maximum waiting time: the process fluctuates at an approximately uniform rate across the entire Markov chain, without spending a disproportionate amount of time in any one state. Gaussian absorption times have also been found in the continuum limit via the linear-noise approximation, which removes state dependence from the noise \cite{hufton2020first}. This approximation is similar to the condition~(\ref{universalityCondition}), which requires the noise amplitude $b_m+d_m$ to vary only mildly across states.  }

 If Eq.~(\ref{universalityCondition}) holds, then $\kappa_n(r, N) \sim c_n(r) f(N)^n  N$, where $f(N) \sim \max_{1\leq i\leq N} (b_i + d_i)^{-1}$. Since these asymptotics hold for $r=r_{\min}$ and $r=r_{\max}$, it follows that $\kappa_n(\{r_j\},N) \sim c_n(\{r_j\}) f(N)^n N$ as well.

\begin{figure}[t]
\includegraphics[width=\linewidth]{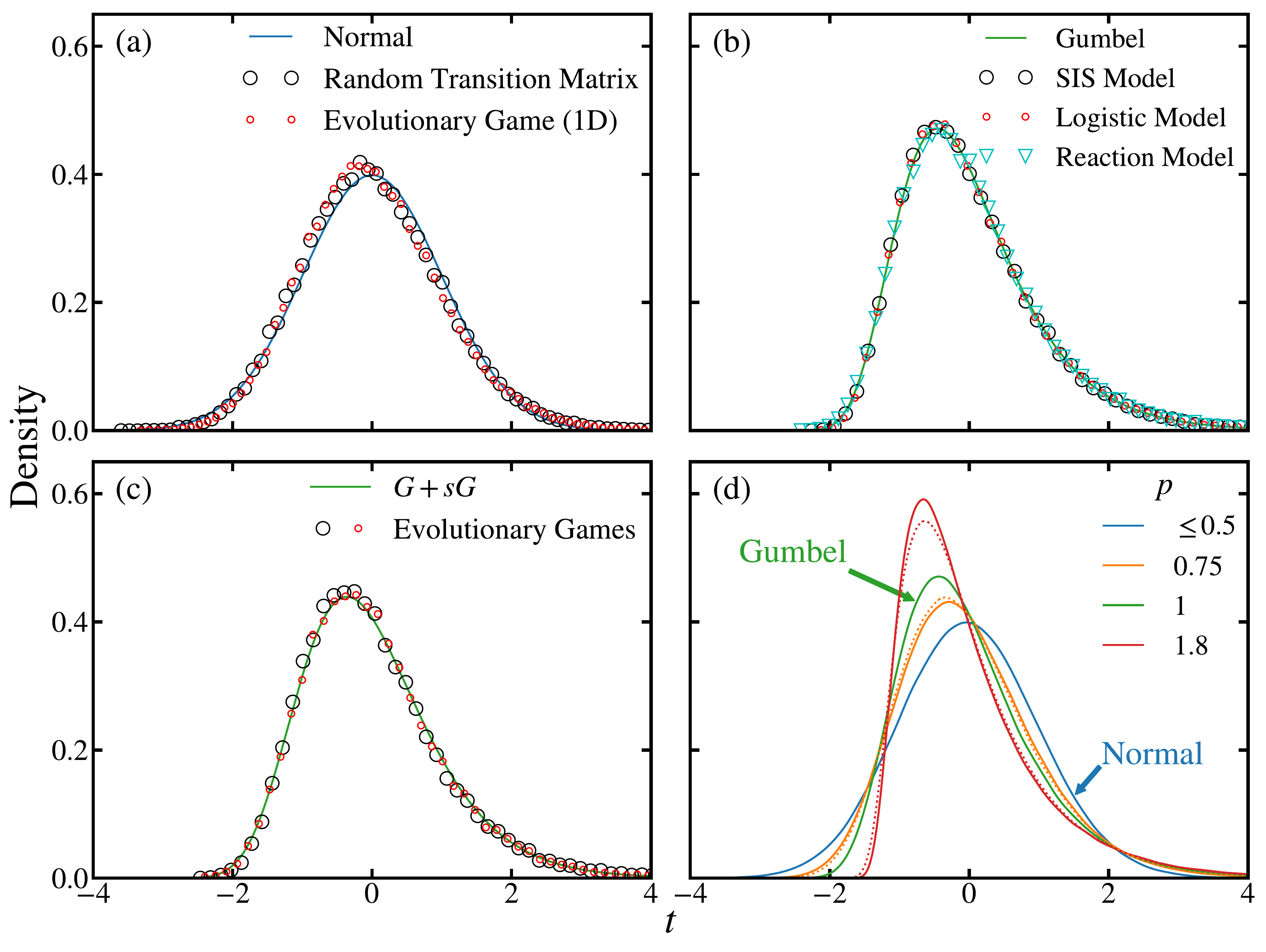} 
	
\caption{\label{fig:allDistributions} {\color{black} Absorption-time distributions for (a)  the random transition matrix model (large black circles) and the evolutionary game on a ring (small red circles), (b) SIS model (large black circles), logistic model (small red circles), and autocatalytic chemical reaction model (cyan triangles), (c) the well-mixed evolutionary game, and (d) the process $b_m = r d_m = r m^p$, for $r=0$ and $p=0.3$  (blue), $p=0.75$ (orange), $p=1$ (green), and $p=1.8$ (red). The $r=0.8$ distributions are indicated by dotted lines (when they differ from the $r=0$ counterparts). See \cite{SM} for models and parameters. We used system sizes  (a-b) $N=500$ and (c-d) $N=1000$ and simulated (a) $5\times 10^4$,  (b-c) $10^5$, and (d) $10^6$ trials to measure the distributions, which have been standardized to have zero mean and unit variance. In (c) the distributions are a convolution of Gumbel distributions with relative weighting $s\approx 0.73$.  Deviations from predicted normal and Gumbel distributions in (a-c) are due to finite system size.}
}

\end{figure}

With the asymptotic form of the cumulants established, we analyze the shape of the distribution using the standardized cumulants $\tilde \kappa_n=  \kappa_n/ \kappa_2^{n/2}$ for $n\geq2$ (which are rescaled so that the variance $\tilde \kappa_2 = 1$). Using the asymptotic form obtained above, we find $\tilde \kappa_n \sim \tilde c_n N^{1-n/2}$. In particular, $\tilde \kappa_n \rightarrow 0$ as $N \rightarrow \infty$ for $n>2$, so that the distribution becomes Gaussian for large $N$ (the cumulants past second order vanish for normal distributions).

For finite $N$, the dominant correction to the normal distribution comes from the non-zero skew $\tilde \kappa_3 \sim \tilde c_3/\sqrt{N}$. The coefficient in this scaling depends on the ratios $r_j$; in the Supplemental Material \cite{SM} we compute a bound on this coefficient, which is useful for estimating the rate of convergence in applications. {The ratio of the standard deviation $\kappa_2^{1/2}$ to the mean $\kappa_1$ also scales like $\kappa_2^{1/2}/\kappa_1 \sim \tilde c_1/ \sqrt{N}$, similar to the skew. As the distribution converges to the Gaussian, the relative width of the distribution narrows at the same rate.}  To summarize, any birth-death Markov chain that satisfies the `flatness' condition, Eq.~(\ref{universalityCondition}), and has an absorbing state toward which the system flows on average ($r_j<1$) will have asymptotically Gaussian distributed absorption times.


Our first example of a Markov chain with normally distributed absorption times is a toy model with \emph{random} transition probabilities. Here we select $b_m+d_m$ uniformly at random between 0.1 and 2 and $r_m$ uniformly at random between $0$ and $0.9$, which satisfies the conditions described above. This example shows that the transition rates need not be smooth in $m$; systems with disordered transition rates still belong to this universality class.

Next we study evolutionary game dynamics on a one-dimensional ring \cite{ohtsuki2006evolutionary, altrock2017evolutionary}. {\color{black} Mutant and wild-type individuals compete via the following dynamics: an individual is chosen randomly, proportional to its (frequency dependent) fitness. The selected individual gives birth to an offspring of the same type, which in turn replaces a random neighbor. The model runs until the mutation spreads to the entire population. }

Figure~\ref{fig:allDistributions}(a) shows simulation results for the random transition system and the evolutionary game. Both display the expected normal distribution. Interestingly, for the evolutionary game, the normal distribution appears for a wide range of parameters, while the mean absorption time and absorption probability depend more intricately on parameters \cite{ohtsuki2006evolutionary, altrock2017evolutionary}.

Gumbel distributions, known for their role in extreme value theory \cite{gumbel1954statistical}, also arise generically in absorption processes. This second universality class is closely related to the `coupon collector' problem in probability theory, which asks the following: if there are $N$ distinct coupons and we are given a random one (with replacement) at each time step, how long does it take to collect all $N$ coupons? The collection process displays a characteristic slowdown: when nearly all coupons have been collected, it takes a long time to acquire the final few because duplicates keep getting selected. Erd\H{o}s and R\'{e}nyi showed that for large $N$ the time to complete the collection follows a Gumbel distribution \cite{erdos1961classical}. 


The coupon collector problem can be modeled using Markov chains. Let $m$ be the number of coupons missing from the collection of $N$ total coupons. The probability of obtaining a new coupon (thereby decreasing $m$) is $m/N$ and the number of missing coupons never increases. Thus, the coupon collection process is described by birth-death dynamics with $b_m=0$ and $d_m = m/N$. The linear decay of the transition probability $d_m$ near the absorbing boundary is the key feature that gives rise to the characteristic slowdown. {\color{black} For this case the cumulants can be computed exactly, $\tilde \kappa_n = (n-1)! \zeta(n)/\zeta(2n)^{n/2}$, and match those for a Gumbel distribution.  Similar to the Gaussian class above, we find that the Gumbel distribution is preserved for non-zero $r_m<1$ and nonlinear transition rates as long as the linear decay is dominant near $0$.} Specifically, if $b_m+d_m = f(N) m [1 + \mathcal{O}(m/N)]$, with $b_{\alpha N} + d_{\alpha N}$ of order at least $\mathcal{O}(N f(N))$ for any $0<\alpha\leq1$, and if $r_{m} = r + \mathcal{O}(m/N)$ for large $N$, then the absorption-time distribution is asymptotically Gumbel
\footnote{{\color{black}More generally it is sufficient to have $b_m+d_m = f(N) m [1 + \mathcal{O}(m/g(N)]$ for any function $g(N)$ that diverges for $N\to \infty$. If $g(N)$ grows linearly or sublinearly, deviations in the cumulants scale like $\delta \tilde \kappa_n \sim C_n [\ln g(N)]/g(N)$. Otherwise, $\delta \tilde \kappa_n \sim C_n/N$}}.

\begin{figure}[t]
	\includegraphics[width=\linewidth]{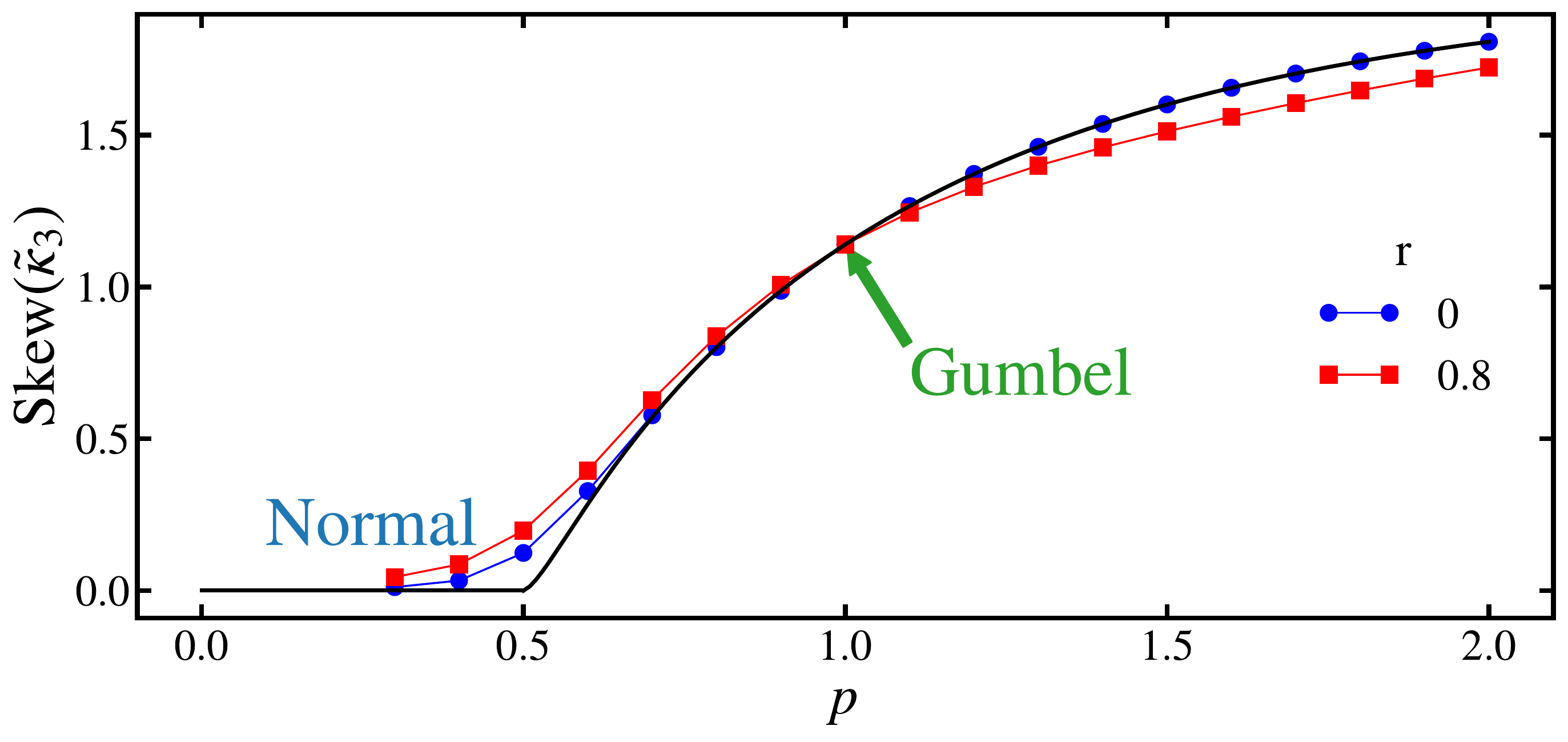}\,
\caption{\label{fig:skewPhaseTransition} {\color{black} Absorption-time skew for the process $b_m = r d_m = r m^p$ with $r=0$ (blue circles) and $r=0.8$ (red squares), plotted as a function of the power-law exponent $p$. Skews were numerically computed for $N=10^5$ using the recurrence relation approach described in Ref.~\cite{hathcock2019fitness}. The black line shows the asymptotic skew $2 \zeta(3 p)/\zeta(2 p)^{3/2}$ for $r=0$. The curves cross at $p=1$ where the distribution is Gumbel, independent of $r$. For $p\leq0.5$ the skew approaches zero and the distribution is Gaussian. The numerical skew is slightly larger than expected for $p\lesssim0.6$ due to finite size effects.}
}
\end{figure}

By bounding the cumulants~(\ref{exactCumulants}), we show \cite{SM} their leading order behavior for $N\gg1$ is dominated by the states near 0, where the approximations $b_m+d_m \approx f(N) m$ and $r_m \approx r$ become asymptotically exact, so that
\begin{equation}\label{gumbelApprox}
\kappa_n(\{r_j\}, N) \sim \frac{1}{f(N)^n} \sum_{1\leq i_1 \leq i_2 \leq \dots \leq i_n\leq N} \frac{w_{i_1i_2 \cdots i_n}^n(r)}{i_1 i_2 \cdots i_n}.
\end{equation}
The factors $f(N)^n$ set the timescale of the process but do not affect the shape of the distribution (they cancel in $\tilde \kappa_n=  \kappa_n/ \kappa_2^{n/2}$). Thus, we have shown that the cumulants are asymptotic to those for a process with $b_m+d_m = m$ and $b_m/d_m =r$. The absorption-time distribution for this process can be computed exactly (see Ref.~\cite[][Appendix B]{azaele2016statistical}) and approaches a Gumbel distribution as $N \to \infty$ \cite{SM}. Therefore, any system with transition rates vanishing linearly and ratios $r_j$ that approach a constant near the absorbing boundary will fall into the Gumbel universality class. 

{As in the Gaussian class, the relative width of the Gumbel distributions becomes small for $N\gg1$. In this case, however, the standard deviation-to-mean ratio scales like $\kappa_2^{1/2}/\kappa_1 \sim C_1/\ln N$. On the other hand, the deviations from the Gumbel cumulants decay like $\delta \tilde \kappa_n =  \tilde \kappa_n -  \tilde \kappa_n^\mathrm{Gumbel} \sim C_n N^{-1} \ln N$ (see Supplemental Material, Section S3.A \cite{SM} and \cite{Note1}). Thus the distribution narrows very slowly compared to the convergence to the Gumbel shape. Therefore, in applications we expect to see the Gumbel distribution appear before the fluctuations become negligible.}

Finally, if the transition rates vanish near the initial condition $N$, scaling like $b_m + d_m = \tilde{f}(N) (N-m) + \mathcal{O}((N-m)^2)$, there will be another coupon-collection slowdown at the beginning of the process. An identical analysis to that above shows that the contributions from the two coupon collection regions simply add together to give the cumulants. The resulting absorption-time distribution is therefore a convolution of two Gumbels, with one weighted by $s = \lim_{N\rightarrow \infty} f(N)/\tilde{f}(N)$.

To illustrate the Gumbel universality class we use the susceptible-infected-susceptible (SIS) model of epidemiology \cite{jacquez1993stochastic}, the logistic model from ecology \cite{grasman1997local}, and an autocatalytic chemical reaction model \cite{doering2007asymptotics, van1992stochastic} (details in Supplemental Material~\cite{SM}). In each case the transition rates decrease linearly near the absorbing state. For example, in the SIS model, $b_m = \Lambda m (1-m/N)$ and $d_m=m$, where $\Lambda$ is the infection rate. 

Our simulations show that these models each have the expected Gumbel distribution [Fig.~\ref{fig:allDistributions}(b)]. The distribution is also insensitive to parameter choices (e.g., a Gumbel appears in the SIS model for any $\Lambda<1$).


\begin{figure}[t]
	\includegraphics[width=\linewidth]{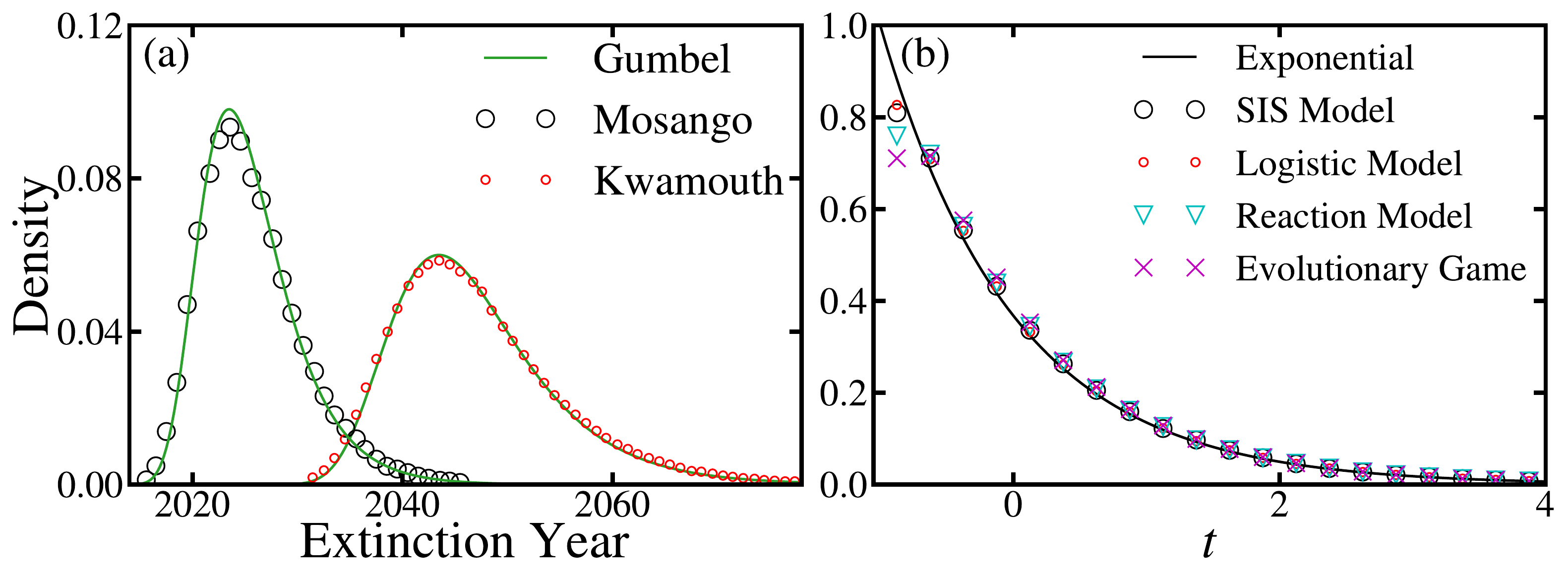}\,
\caption{\label{fig:sleeping_sickness_exponentials} Generalizations to high-dimensional models and Markov chains with internal sinks. (a) Extinction-time distributions for sleeping sickness predicted using a 17-dimensional compartmental model that was fit to case data from the Mosango (large black circles) and Kwamouth (small red circles) regions of the Democratic Republic of Congo (data from Ref.~\cite{aliee2020estimating}). Mean extinction times (measured from 2016) are approximately 9.5 and 31 years for the Mosango and Kwamouth regions respectively, with standard deviations of 4.8 and 7.9 years. Disease eradication times approximately follow a Gumbel distribution (fit using the mean and variance). {\color{black} (b) Simulations of the SIS, logistic, reaction, and well-mixed evolutionary game models have exponential absorption-time distributions (standardized to zero mean and unit variance) if parameters are chosen so that the dynamics have an internal sink state. For each case, we used $N=50$ and simulated $10^6$ trials. See \cite{SM} for model details and parameters.}}
\end{figure}

If we study the aforementioned evolutionary game in a well-mixed population, the transition rates vanish linearly as $m\rightarrow0$ and $m \rightarrow N$ \cite{ashcroft2016statistical, SM}. As discussed above, we expect a convolution of Gumbel distributions with relative weighting $s$ given by the ratio of the linear coefficients at these two boundaries. Figure~\ref{fig:allDistributions}(c) shows that this prediction is borne out in simulations. 

{\color{black}  In addition to Gumbel and Gaussian classes, other absorption-time distributions arise if the transition rates have power-law decay: $b_m+d_m = f(N) m^p [1+ \mathcal{O}(m/N)]$. For $p<1/2$, the decay is sufficiently slow that the normal distribution is maintained: the system still fluctuates at an approximately uniform rate across states. On the other hand, if $p>1/2$ we find a generalized coupon collection phenomenon giving rise to a family of skewed distributions. Slowdown near the boundary dominates the absorption process and the distribution is asymptotic to that for the minimal model $b_m = r d_m = r m^p$ \cite{SM}. When $r=0$ the cumulants can be computed analytically: $\tilde \kappa_n = (n-1)! \zeta(n p)/\zeta(2p)^{n/2}$ \cite{ottino2017evolutionary, ottino2017takeover}. Figure~\ref{fig:allDistributions}(d) shows the resulting distributions for a few values of $p$. Interestingly for $p\neq1$, the shape of the distribution depends subtly on $r$. Figure~\ref{fig:skewPhaseTransition} shows the skew of these distributions as a function of $p$, elucidating the transition from normal distributions to the skewed family.}

{\color{black} Beyond simple one-dimensional Markov processes,} the eradication-time distributions for African sleeping sickness predicted by a 17-dimensional data-driven model \cite{aliee2020estimating} closely resemble the Gumbel [Fig.~\ref{fig:sleeping_sickness_exponentials}(a)]. This result suggests that the Gumbel distribution is also generic in higher dimensions if the dynamics collapse onto a  one-dimensional slow manifold near absorption. {\color{black} Crucially, although the distributions have converged to the Gumbel shape, the fluctuations  still matter: the probable extinction times span  years. The ratio between the standard deviation and the mean is approximately $0.5$ and $0.25$ for the Mosango and Kwamouth regions respectively.} {\color{black} Similar results hold for a variety of high-dimensional systems. Their dynamics are accurately approximated by birth-death processes with transition rates that vanish as a power-law $m^p$ near the boundary. Examples include evolutionary dynamics on $D$-dimensional lattices ($p = 1-1/D$) and complex networks \cite{hajihashemi2019fixation, ottino2017evolutionary, ottino2017takeover} as well as epidemics on networks \cite{dilauro2020network}.   }

In this Letter we have characterized universality classes for absorption times in birth-death Markov chains. While our results are formulated in terms of the transition rates $b_i$ and $d_i$, we can also connect the shape of the absorption-time distribution to the spectrum of the transition matrix. 
Discussion and derivation of these results are provided in the Supplemental Material, sections S2.B and S3.C \cite{SM}.
Future work might focus on characterizing additional universality classes beyond those studied here. For example, simulations [Fig.~\ref{fig:sleeping_sickness_exponentials}(b)] show that exponential absorption-time distributions arise frequently in systems with an internal sink state, toward which transitions are more likely \cite{yahalom2019comprehensive}.  {\color{black} The emergence of the exponential distribution makes sense intuitively: the system quickly settles into a quasiequilibrium mode around the sink, whose slow exponential decay dominates the absorption process \cite{collet2013quasi}. To our knowledge, however, there is no rigorous classification of this case.  It would also be fascinating to investigate whether there is a universal crossover between different members of our family of absorption-time distributions. For example, how do the distributions change if the transition rates have mixed decay $m^p + \epsilon m^q$? Understanding the crossover scaling between these cases will enable the classification for an even broader class of extinction-prone Markov chains. }

\vspace{3mm}
\begin{acknowledgments}
We thank David A. Kessler and Nadav Shnerb for helpful comments regarding the Gumbel classification and connections to extreme value theory. This work was supported by an NSF Graduate Research Fellowship, grant No. DGE-1650441 to D.H.
\end{acknowledgments}


\bibliography{references}

\end{document}


\title{Supplemental Material for ``Asymptotic absorption-time distributions in extinction-prone Markov processes"}

\author{David Hathcock}
\affiliation{Department of Physics,  Cornell University, Ithaca, New York 14853, USA}
\author{Steven H. Strogatz}
\affiliation{Department of Mathematics,  Cornell University, Ithaca, New York 14853, USA}

\date{\today}

\maketitle

\onecolumngrid

This Supplemental Material provides rigorous mathematical derivations supporting the results discussed in the main text and gives details of the various example models used to demonstrate the Gaussian and Gumbel classes. {\color{black} Section~\ref{robustness} provides discussion of additional cases (with different initial or boundary conditions) where our results still apply.} Section~\ref{cumulantDerivation} gives a detailed derivation of the absorption-time cumulants (Eq.~(2) in the main text), including the weighting factors $w^n$ and their properties. Sections~\ref{normalAsymptotics} and \ref{gumbelAsymptotics} provide the details of the asymptotic analysis used to characterize Markov processes with normal and Gumbel absorption-time distributions respectively and discuss the leading corrections and transition matrix spectrum for each case. {\color{black} Section~\ref{skewedFamily} shows the classification of processes with transition rates that decay as a power-law with exponent $p$ near the boundary. We show this power-law decay leads to Gaussian distributions ($p\leq1/2$), or a skewed family of distributions ($p>1/2$)}. Finally, in  Section~\ref{exampleModels} we describe the dynamics for the evolutionary game, the susceptible-infectious-susceptible model, the logistic model, and the autocatalytic chemical reaction model. For each case we derive the Markov transition probabilities and explain how the model falls into one of our universality classes. 

{\color{black} \section{Distributions are robust to changes in initial and boundary conditions}\label{robustness}
In the main text, we specialize to Markov chains with a finite state space of size $N$, a reflecting upper boundary, and initial condition at the maximal state $p_m(0) = \delta_{m,N}$. Our asymptotic absorption-time distributions, however, should be robust to changes in initial and boundary conditions. Because the dynamics are extinction-prone, the system quickly progresses toward the absorbing state, spending  negligible time near the reflecting boundary. Therefore, if the initial condition $m_0$ is sufficiently large ($m_0\sim N $ for large $N$), corrections due to variation in the initial condition will be sub-dominant as $N \rightarrow \infty$. By the same argument, we expect the same asymptotic distributions to occur for infinite systems with free boundary conditions and no maximal state $N$, but large initial condition. On the other hand, if the upper boundary is absorbing, our result describes the absorption-time distribution, given that the absorbing state at $0$ is reached (i.e. if we ignore all trajectories that are absorbed at the upper boundary) \cite{hathcock2019fitness}. Finally, our results can also be used to determine the first-passage-time distribution to an arbitrary state $m$, since the first-passage problem can be solved by making the target state absorbing \cite{van1992stochastic}.

}

\section{Absorption-time cumulants and weighting factor properties}\label{cumulantDerivation}

\subsection{Derivation of the absorption-time cumulants}

In this section we derive a general formula for the absorption-time cumulants, shown as Eq.~(2) in the main text. This derivation follows Ref.~\cite{hathcock2019fitness}, but we generalize to Markov chains where the ratio $r_m = b_m/d_m$ is non-constant. We start from the master equation, Eq.~(1) in the main text, and restrict our attention to the transient (non-absorbing) states, $m>0$, since these determine the time it takes to reach absorption. The master equation for these states can be expressed as $\dot{\mathbf{p}}(t) = \Omega \cdot \mathbf{p}(t)$, where $\Omega$ is the transient transition matrix with elements 
\begin{equation}
\Omega_{mn} = b_n \delta_{m,n+1} + d_n \delta_{m,n-1} - (b_n + d_n) \delta_{m,n} 
\end{equation}
for $m,n = 1, \dots, N$ and $\mathbf{p}(t)$ is the vector of transient state occupancy probabilities. 

The entire first-passage process can be characterized in terms of the transition matrix $\Omega$. In fact, the first-passage distribution $p(t)$ can be written in terms of an element of the matrix exponential, $p(t) = d_1 [\exp(\Omega t )] _{1,N}$ and the moments of $T$ are
\begin{equation}\label{moments}
\mu_n \coloneqq E[T^n] = (-1)^n n! \mathbf{1} \Omega^{-n} \mathbf{p}(0),
\end{equation}
where $\mathbf{1}$ is a row vector containing all 1's and $E$ denotes expected value. 

As discussed in the main text, to proceed it is useful to introduce the following decomposition of the transition matrix: $\Omega = \Omega_{RW} D$, where $D$ is a diagonal matrix $D_{mm} = b_m + d_m$ and 
\begin{equation}
[\Omega_{RW}]_{mn} = \frac{r_n}{1+r_n} \delta_{m,n+1} + \frac{1}{1+r_n} \delta_{m,n-1} -  \delta_{m,n},
\end{equation}
with $r_n = b_n/d_n$. The rates $D_{mm}$ determine how long the system waits in state $m$ before taking a step and $r_m$ is the relative probability of stepping forward versus backward along the chain. Defining $V = -\Omega_{RW}^{-1}$, the elements $V_{ij}$ are the average number of visits to state $i$ before absorption starting from an initial state $j$.

With the above decomposition we can easily invert the transition matrix, 
\begin{equation}
    [-\Omega^{-1}]_{ij} = \frac{V_{ij}}{b_i+d_i},
\end{equation}
where visit numbers $V_{ij}$ are given by
\begin{equation}\label{visitNumbers}
V_{ij} = \left (1+r_i\right)  \sum_{n=1}^{\min(i,j)} \prod_{m=n}^{i-1} r_m.
\end{equation}
Then, using Eq.~(\ref{moments}) the moments can be expressed as
\begin{equation}\label{momentsSeries}
    \mu_n = n! \sum_{i_1,i_2,\dots i_n=1}^N \frac{V_{i_1i_2} V_{i_2i_3} \cdots V_{i_{n-1}i_n} V_{i_n N}}{(b_{i_1} + d_{i_1})(b_{i_2} + d_{i_2}) \cdots (b_{i_n} + d_{i_n})}.
\end{equation}
To compute the cumulants, we use the standard conversion formulas: $\kappa_1 = \mu_1$, $\kappa_2 = \mu_2 - \mu_1^2$, $\kappa_3 = \mu_3 - 3 \mu_2 \mu_1 + 2 \mu_1^3$, and so on. Since the relation between cumulants and moments is polynomial, if we collect terms with common denominators it follows that the cumulants have the form quoted in the main text,
\begin{equation}\label{exactCumulants}
\color{black}    \kappa_n(\{r_j\}, N) = \sum_{1\leq i_1 \leq i_2 \leq \dots \leq i_n \leq N} \frac{w_{i_1i_2 \cdots i_n}^n(\{r_j\})}{(b_{i_1}+d_{i_1})\cdots(b_{i_n}+d_{i_n})}.
\end{equation}
where the weights $w^n$ depend on the visit numbers $V_{ij}$ (and hence are functions of only the ratios $\{r_j\}$). {\color{black} Note that we sum over $i_1\leq i_2\leq \cdots \leq i_n$, so that each product in the denominator of Eq.~(\ref{exactCumulants}) appears exactly once. The weights are determined using Eq.~(\ref{momentsSeries}) and the moment-cumulant conversion formulas.} For example, the second and third cumulants are
\begin{align}
    \kappa_2 &= \sum_{i,j=1}^N \frac{2 V_{ij}V_{jN}-V_{iN}V_{jN}}{(b_i+d_i)(b_j+d_j)} \\
    \kappa_3 &= \sum_{i,j,k=1}^N \frac{6 V_{ij} V_{jk} V_{kN}-6 V_{ij}V_{jN}V_{kN}+2 V_{iN} V_{jN} V_{kN}}{(b_i+d_i)(b_j+d_j)(b_k+d_k)}.
\end{align}
From here we can read off the weights $w^n$: they are simply the numerators in the above expressions, summed over distinct permutations of the indices (since these terms have the same denominators). Carrying out the sum we obtain,
\begin{align}
    w^2_{ij} &=  \sum_{\sigma \in \Pi_2} 2 V_{\sigma_1 \sigma_2}V_{\sigma_2 N}-V_{\sigma_1 N}V_{\sigma_2 N} \\
    w^3_{ijk} &= \sum_{\sigma \in \Pi_3} 6 V_{\sigma_1 \sigma_2} V_{\sigma_2 \sigma_3} V_{\sigma_3 N}-6 V_{\sigma_1 \sigma_2}V_{\sigma_2N}V_{\sigma_3N}+2 V_{\sigma_1N} V_{\sigma_2N} V_{\sigma_3N},
\end{align}
where $\Pi_2$ is the set of distinct permutations of indices $\{i,j\}$ and $\Pi_3$ is the set of distinct permutations of $\{i,j,k\}$.

\subsection{Properties of the weighting factors $w^n$}
In the main text, we noted that the weights $w^n$ have a few convenient properties. In particular, they are positive, $w_{i_1i_2 \cdots i_n}^n(\{r_j\}) \geq 0$ and increasing functions of each of the $r_j$. To show these properties, we use the fact that $V_{ii} = V_{ij}$ for any $i<j$. This is easy to see from Eq.~(\ref{visitNumbers}), but also has an intuitive physical interpretation. Since the system is eventually absorbed at the boundary state $0$, if it starts from a state $j>i$ it must visit $i$ before absorption. After the first visit, the statistics of the random walk are identical to a walk initialized in state $i$. Using this property the sum over permutations above dramatically simplifies. For $i < j$ we have
\begin{align}\label{weights2}
    w^2_{ij} &= (2 V_{ij}V_{jN}-V_{iN}V_{jN} + 2 V_{ji}V_{iN} - V_{jN}V_{iN})\\
    &= 2 V_{ji}V_{iN}. \nonumber
\end{align}
Similarly, after simplification we find 
\begin{align}
    w^3_{ijk} &= 3!\, (V_{kj} V_{ji}V_{iN} + V_{ki}V_{iN} V_{jN} )\\
    w^4_{ijkl} &= 4! \, (V_{lk}V_{kj}V_{ji}V_{iN} + V_{lk} V_{ki} V_{iN} V_{jN} + V_{lj}V_{ji}V_{iN} V_{kN} \\
    &\quad \quad+ V_{lj}V_{jN}V_{ki}V_{iN} + V_{li}V_{iN}V_{kj}V_{jN}+V_{li}V_{iN}V_{kN}V_{iN} )\nonumber
\end{align}
when $i< j < k < l$.  When some indices are identical, these results still hold, but they must be divided by the number of permutations of the identical indices, e.g. $w_{ii}^2 = V_{ii}V_{iN}$ (notice this differs from Eq.~(\ref{weights2}) by a factor of 2). The important feature of these expressions is that they are \emph{positive} sums of products of the visit numbers $V_{ij}$. {\color{black} We conjecture that the weights at every order can also be written as a positive sums of products of the visit numbers (though we omit the expressions here, we have checked this is true up to order $n=6$). If this is the case, it immediately follows that the weights $w^n$ are positive and  increasing functions of each $r_j$ because the visit numbers, Eq.~(\ref{visitNumbers}), themselves also have these properties.}

\section{Asymptotic analysis for the Gaussian Universality Class}\label{normalAsymptotics}

\subsection{Cumulant bounds}
 
To estimate the asymptotics of the cumulants we start from Eq.~(\ref{exactCumulants}) derived above. Since the weights $w^n$ are increasing functions of the $r_j$, we argued in the main text that 
\begin{equation}\label{maxminBound}
\kappa_n(r_{\min}, N) \leq \kappa_n(\{r_j\}, N) \leq \kappa_n(r_{\max}, N),
\end{equation}
where $r_{\max} = \lim_{N\rightarrow\infty} \max_{1<j<N} r_j$ and $r_{\min} = \lim_{N\rightarrow\infty} \min_{1<j<N} r_j$. The cumulants  $\kappa_n(r_{\max}, N)$ and $\kappa_n(r_{\min}, N)$ correspond to auxiliary Markov chains where $b_j+d_j$ is unchanged, but $r_j = r_{\max}$ or $r_j=r_{\min}$ respectively.

Following Ref.~\cite{hathcock2019fitness}, we provide asymptotic bounds on $\kappa(r, N)$ that lead to an analytic criterion for the Gaussian universality class. Since the diagonal elements of the weights $w^n$ are greater than 1, we can bound the cumulant $\kappa_n$ from below by a sum of the unweighted diagonal elements $(b_i + d_i)^{-n}$. To bound from above we can take the maximum value of $(b_i + d_i)^{-n}$ times the sum of the weighting factors. The sum over weighting factors $w_{i_1i_2 \cdots i_n}^n(r)$ is precisely the $n^\text{th}$ cumulant for a biased random walk (with $b_i+d_i=1$ and uniform $r$). This sum can be computed exactly using eigenvalues of the transition matrix \cite{hathcock2019fitness}. In particular, the sum is $\mathcal{O}(N)$ for any $n$ as long as $r<1$ and asymptotically can be represented in the integral form given below. Note that $r_{\max}<1$ as long as $r_j<1-\epsilon$ for all $j$ and some $\epsilon>0$: this condition was the first requirement for the Gaussian universality class quoted in the main text. Altogether we have,
\begin{equation}\label{normalBounds}
\begin{split}
\sum_{n=1}^N \frac{1}{(b_i + d_i)^n} \leq \kappa_n(r, N) \leq & \left( \max_{1\leq i \leq N} \frac{1}{b_i + d_i} \right)^n \times \frac{N}{\pi} \int_0^\pi \frac{(n-1)!}{(1-2 \sqrt{r}/(1+r) \cos x)^n} \mathrm{d}x \\
 = &\left( \max_{1\leq i \leq N} \frac{1}{b_i + d_i} \right)^n \times \mathcal{O}(N).
\end{split}
\end{equation}

We can now read off the second condition for the Gaussian universality class. To nail down the asymptotics of $\kappa_n(r,N)$ we want the upper and lower bounds in Eq.~(\ref{normalBounds}) to have the same scaling for large $N$. Specifically, we require
\begin{equation}\label{universalityCondition}
\frac{1}{N} \sum_{i=1}^N \frac{1}{(b_i + d_i)^n}  \sim c_n \left( \max_{1\leq i \leq N} \frac{1}{b_i + d_i} \right)^n,
\end{equation}
for some $N$-independent constant $c_n$. Setting $n=1$ in this equation leads to the condition Eq.~(3) quoted in the main text. We can make this simplification because when Eq.~(\ref{universalityCondition}) is satisfied for $n=1$, it is also satisfied for $n>1$. To see this fact, first note that  $\langle (b_i + d_i)^{-n} \rangle < \max_i (b_i + d_i)^{-n}$ trivially. Furthermore, we can write the left hand side of Eq.~(\ref{universalityCondition}) as $N^{-1} || (b+d)^{-1}||_p^p$, where $|| \cdot ||_p$ is the $p$-norm and $(b+d)^{-1}$ is the vector containing elements $(b_i + d_i)^{-1}$. Using $p$-norm inequalities, we have  $N^{-1} || (b+d)^{-1}||_1 < N^{-1/n} || (b+d)^{-1}||_n$ for $n>1$. Then if $ c \cdot \max_i (b_i + d_i)^{-1} < \langle (b_i + d_i)^{-1} \rangle$ as $N \to \infty$ for some constant $c$, it follows that $ c^n \max_i (b_i + d_i)^{-n} < \langle (b_i + d_i)^{-n} \rangle$ in this limit as well. Thus, it is sufficient to check Eq.~(\ref{universalityCondition}) holds for $n=1$,  since this implies the condition holds for all  $n>1$.

As discussed in the main text, the condition Eq.~(\ref{universalityCondition}) can be interpreted as the waiting times being `flat' in the following sense: all (or at least a significant fraction) of the $(b_i+d_i)^{-1}$ are the same order as their maximum value. If this condition holds, then for large $N$ we have that $\kappa_n(r,N) \sim c_n(r) f(N)^n N$ where $f(N) \sim \max_{1\leq i\leq N} (b_i + d_i)^{-1}$ as $N\rightarrow \infty$. Since these asymptotics hold for both $r=r_{\min}$ and $r=r_{\max}$, it follows from Eq.~(\ref{maxminBound}) that $\kappa_n(\{r_j\},N) \sim  c_n(\{r_j\}) f(N)^n N$ as well, possibly with a different constant $c_n(r_\mathrm{min})<c_n(\{r_j\}) <c_n(r_{\max})$. These asymptotics imply that the higher-order cumulants are dominated by the variance and hence the distribution looks normal for large $N$, i.e. the standardized cumulants $\tilde \kappa_n = \kappa_n/\kappa_2^{n/2} \to 0$ as $N\to\infty$.

\subsection{Leading correction to the Gaussian}
The leading correction to the Gaussian distribution for finite $N$ comes from the skew, $\tilde \kappa_3 = \tilde c_3/\sqrt{N}$. Here we will give a bound on the magnitude of the skew, that can be used to predict when finite systems will have a nearly Gaussian absorption-time distribution. First, define 
\begin{equation}
    K_2 = \lim_{N\to\infty}\frac{1}{N f(N)^2}\sum_{i=1}^N \frac{1}{(b_i + d_i)^2},
\end{equation}
where $f(N) \sim \max_{1\leq i\leq N} (b_i + d_i)^{-1}$ as $N\rightarrow \infty$ as above. Then from Eq.~(\ref{normalBounds}) that $\kappa_2 \geq K_2 N f(N)^2$. Evaluating the integral in Eq.~(\ref{normalBounds}) we have $\kappa_3 \leq 2 f(N)^3 N (r_{\max}+1)^3(r_{\max}^2+4r_{\max}+1)/(1-r_{\max})^5$. Putting these together,
\begin{equation}\label{skewbound}
\tilde \kappa_3 \leq  \frac{2 (r_{\max}+1)^3(r_{\max}^2+4r_{\max}+1)}{(1-r_{\max})^5 K_2^{3/2}}\frac{1}{\sqrt{N}}.
\end{equation}
The convergence is slowest (i.e. the coefficient of $1/\sqrt{N}$ is large), when the conditions for the universality class are pushed to their limits: if the system is barely extinction-prone, $r_{\max} \approx 1$, or the waiting times are not very uniform, $K_2 \ll 1$ (the sum in Eq.~(\ref{universalityCondition}) is nearly dominated by the maximal term). Finally, we note that this is a rough upper bound; in many cases the convergence is much faster, e.g., if only a few $r_j\approx1$ but the rest are very small. Replacing $r_{\max}$ with the average $r_j$ in Eq.~(\ref{skewbound}) may often give a better estimate of the actual skew for a given system, even if it does not give a strict upper bound.

\subsection{Transition matrix spectrum}
As noted in the main text, we can also connect the spectrum of the transition matrix to the Gaussian universality class. Specifically, if Eq.~(\ref{universalityCondition}) is satisfied with $b_i+d_i$ replaced by the eigenvalues $\lambda_i$ of the negative transition matrix $-\Omega$, the absorption-time distribution will be Gaussian. To show this, we use the spectral representation of the absorption-time cumulants \cite{hathcock2019fitness, ashcroft2016statistical},
\begin{equation}\label{eigenCumulants}
    \kappa_n = (n-1)! \sum_{i=1}^{N} \lambda_i^{-n}.
\end{equation}
If Eq.~(\ref{universalityCondition}) is satisfied for the eigenvalues, we have 
\begin{equation}
\sum_{i=1}^N \lambda_i^{-n} \sim c_n \left( \max_{1\leq i \leq N} \lambda_i^{-1} \right)^n,
\end{equation}
Since the left-hand side of this expression is exactly the cumulant $\kappa_n$ (up to the constant $(n-1)!$), it immediately follows that $\kappa_n\sim c_n g(N)^n N$ where $g(N) \sim \max_{1\leq i \leq N} \lambda_i^{-1}$ as $N \rightarrow \infty$. Just as in the main text, this scaling implies that the standardized cumulants vanish for large $N$: $\tilde \kappa_n \sim \tilde c_n N^{1-n/2}$ and the distribution is asymptotically Gaussian. 

More generally, the distribution approaches a Gaussian as long as $\tilde \kappa_n \rightarrow 0$ as $N\to \infty$. This condition with Eq.~(\ref{eigenCumulants}) describes a broader class of eigenvalue spectra that give rise to Gaussian absorption-time distributions. Specifically, we need
\begin{equation}
    \left (\sum_{i=1}^N \lambda_i^{-n} \right) \Bigg/ \left (\sum_{i=1}^N \lambda_i^{-2} \right)^{n/2} \xrightarrow[]{N\to\infty}0.
\end{equation}
While this condition is difficult to interpret, we consider two  examples that illustrate the type of spectra that can give rise to Gaussian absorption-time distributions. First, if $\lambda_m = m^p$, the above condition is satisfied for $p\leq1/2$. This result is related to the emergence of Gaussian distributions for the systems  considered in Section~\hyperref[powerLawGaussian]{S5.B}, which have transition rates that decay as a power-law with $p\leq1/2$. Also, if $\lambda_m = P(m)/Q(m)$ for some polynomials $P$ and $Q$, the condition is satisfied when the degree of $Q$ is greater than that of $P$.

\section{Asymptotic analysis for the Gumbel Universality Class}\label{gumbelAsymptotics}

\subsection{Cumulant bounds}\label{gumbelBounds}

For the Gumbel universality class we require $b_m+d_m = f(N) m [1+ \mathcal{O}(m/N)]$, $b_{\alpha N} + d_{\alpha N}$ be of order at least $\mathcal{O}(N f(N))$ for any $0<\alpha<1$, and $r_m = r + \mathcal{O}(m/N)$ for large $N$. These properties are sufficient to guarantee that the absorption-time cumulants are asymptotic to those for an exactly solvable canonical model (for which the above equalities hold exactly, not just to leading order). Following Ref.~\cite{hathcock2019fitness}, we restrict two of the indices in Eq.~(\ref{exactCumulants}) to be $\mathcal{O}(N)$ away from the absorbing state, $\alpha N \leq i_{n-1} \leq i_n \leq N$. With this restriction we can bound the sums,
\begin{equation}\label{bound1}
\sum_{\substack{1\leq i_1 \leq i_2 \leq \dots \leq i_{n-1} \mathstrut \\ \mathstrut \alpha N \leq i_{n-1} \leq i_n \leq N}} \frac{w_{i_1i_2 \cdots i_n}^n(\{r_j\})}{(b_{i_1}+d_{i_1})\cdots(b_{i_n}+d_{i_n})} \leq 
\frac{1}{f(N)^n N^2} \sum_{1\leq i_1 \leq i_2 \leq \dots \leq i_n \leq N} w_{i_1i_2 \cdots i_n}^n(r).
\end{equation}
In the previous section, we established that the sum over the weighting factors is $\mathcal{O}(N)$, so this portion of the sum is $\mathcal{O}(f(N)^{-n} N^{-1})$. We now consider indices $1<i_1<\alpha N$ and $\alpha N<i_n<N$,
\begin{equation}\label{bound2}
\sum_{\substack{i_1 \leq i_2 \leq \dots \leq i_{n} \mathstrut \\ \mathstrut 1 \leq i_{1} \leq \alpha N \leq i_n \leq N}} \frac{w_{i_1i_2 \cdots i_n}^n(\{r_j\})}{(b_{i_1}+d_{i_1})\cdots(b_{i_n}+d_{i_n})} \leq 
\frac{1}{f(N)^n N} \sum_{\substack{i_1 \leq i_2 \leq \dots \leq i_{n} \mathstrut \\ \mathstrut 1 \leq i_{1} \leq \alpha N \leq i_n \leq N}}w_{i_1i_2 \cdots i_n}^n(r).
\end{equation}
Since the weighting factors decay exponentially away from the diagonal elements, the sum on the right hand side of Eq.~(\ref{bound2}) is $\mathcal{O}(1)$ and this portion of the sum is also $\mathcal{O}(f(N)^{-n} N^{-1})$.

Since the same bounds also apply for any other pair of the indices, the only remaining portion of the cumulant sum Eq.~(\ref{exactCumulants}) is that where all indices are near $0$. Here the approximations that $b_m+d_m$ is linear and $r_m$ is constant become asymptotically exact so that,
\begin{equation}\label{sumAsymptotics}
  \sum_{1\leq i_1 \leq i_2 \leq \dots \leq i_n \leq \alpha N} \frac{w_{i_1i_2 \cdots i_n}^n(\{r_j\})}{(b_{i_1}+d_{i_1})\cdots(b_{i_n}+d_{i_n})} \sim  \frac{1}{f(N)^n}  \sum_{1\leq i_1 \leq i_2 \leq \dots \leq i_n \leq \alpha N} \frac{w_{i_1i_2 \cdots i_n}^n(r)}{i_1 i_2 \cdots i_n}.
 \end{equation}
The right hand side of Eq.~(\ref{sumAsymptotics}) is at least $\mathcal{O}(f(N)^{-n})$ and therefore this region of the cumulant sum dominates asymptotically compared to the $\mathcal{O}(f(N)^{-n} N^{-1})$ terms estimated above. In other words, the absorption process is entirely dominated by the coupon collection behavior near the absorbing state. Furthermore, we can freely extend the upper limit of the sum to $N$ (instead of $\alpha N$) since this will only add subdominant terms. Finally, we obtain the result quoted in the main text,
\begin{equation}\label{gumbelApprox}
\kappa_n(\{r_j\}, N) \sim \frac{1}{f(N)^n}  \sum_{1\leq i_1 \leq i_2 \leq \dots \leq i_n \leq N} \frac{w_{i_1i_2 \cdots i_n}^n(r)}{i_1 i_2 \cdots i_n}.
\end{equation}
Thus, for any Markov chain satisfying the conditions at the beginning of this section, the cumulants are asymptotic to those for the ``canonical model'' with $b_m + d_m = f(N) m$ and $r_m=r$ exactly. In Section~\hyperref[canonical]{S4.C} we show this model has an asymptotically Gumbel absorption-time distribution.

\subsection{Leading correction to the Gumbel}
{\color{black}
The leading correction $\delta \kappa_2$ to the standard deviation comes from the quadratic term in the transition rates, $b_m + d_m \approx f(N) m (1+m/N)$. Plugging this into Eq.~(\ref{gumbelApprox}) for one set of rates in the denominator and using the partial fraction decomposition $1/i (j+j^2/N) = 1/ij + 1/i(j+N)$ leads to 
\begin{equation}
\begin{split}
\delta \kappa_2 &\sim \frac{1}{f(N)^2 }\sum_{1\leq i \leq j \leq \alpha N} \frac{w^2_{i,j}(r)}{i (j+N)} \\
&=  \frac{1}{f(N)^2 }\sum_{i = 1}^{\alpha N} \frac{(1+r)^2(1-r^i)^2}{(1-r)^2 i (i+N)} + 2 \sum_{i=1}^{j-1}\sum_{j=1}^{\alpha N} \frac{r^{j-i}(1+r)^2(1-r^i)^2}{(1-r)^2 i (j+N)}
\end{split}
\end{equation}
where in the second line we make use of the fact that $r_m \approx r$ is approximately constant to write the explicit expression for $w^2_{i,j}$ (obtained from Eqs.~(\ref{visitNumbers}) and (\ref{weights2})). The sums in the second line can be evaluated explicitly in terms of special functions, including harmonic numbers and the Lerch transcendent. The first sum is asymptotically dominant, leading to $\delta\kappa_2 \sim f(N)^{-2} N^{-1} \ln (N)$. More generally, the asymptotics above hold if $b_m + d_m \approx f(N) m (1+m/g(N))$ as long as the function $g(N) \to \infty $ as $N\to \infty$. An analogous calculation shows that for this case $\delta\kappa_2 \sim [\ln g(N)]/g(N)f(N)^2$. If $g(N)$ grows superlinearly, however, $[\ln g(N)]/g(N)f(N)^2$ is dominated by the corrections due to Eq.~(\ref{bound1}) and (\ref{bound2}) computed above, leading to $\delta\kappa_2 \sim f(N)^{-2} N^{-1} $.

The higher-order cumulants can be analyzed in similar fashion. Since the weights decay exponentially away from the diagonal, the terms with $i_1=i_2=\cdots=i_n \equiv i$ is asymptotically dominant. For these elements $w_{i_1i_2 \cdots i_n}^n(\{r_j\}) = (n-1)! V_{ii}^n$ and it is straightforward to show that $\delta\kappa_n \sim N^{-1} f(N)^{-n}$. for the standardized cumulants $\tilde \kappa_n = \kappa_n/\kappa_2^{3/2}$, the factors of $f(N)$ in the asymptotics cancel and we are left with $\mathcal{O}(N^{-1} \ln N)$ corrections from the standard deviation. In other words, the deviations from the Gumbel cumulants scale like $\delta \tilde \kappa_n = \tilde \kappa_n - \tilde \kappa_n^\mathrm{Gumbel}~\sim C_n N^{-1} \ln N$ for large $N$. For the more general case, where the quadratic term in the rates is suppressed by $g(N)$, the scaling is $\delta \tilde \kappa_n \sim C_n  [\ln g(N)]/g(N)$ for sublinear $g(N)$ and $\kappa_n \sim C_n/N$ otherwise.}

\subsection{Large-$N$ limit for the canonical model}\label{canonical}

The canonical Markov model with coupon-collection behavior has $b_m + d_m = f(N) m$ and $r_m=r$. As noted in the main text, $f(N)$ simply sets the time scale for the process and does not affect the shape of the absorption-time distribution. Therefore, for convenience, we will rescale time $t\rightarrow t (r+1)/f(N)$ so that $b_m = r m$ and $d_m = m$. For this system, the absorption-time distribution $p(t)$ has been computed exactly using generating functions \cite[][Appendix B]{azaele2016statistical},
\begin{equation}
p(t) = \frac{N e^{\nu t} \nu^2}{(e^{\nu t}-1)^2 (1+ \frac{\nu}{e^{\nu t} -1})^{N+1}}
\end{equation}
where $\nu = 1-r$. To derive the asymptotic form of the distribution we standardize to zero mean and unit variance. The standardized distribution is simply $\sigma  p(\sigma t + \mu)$, where $\mu \sim  (\ln N + \ln \nu + \gamma)/\nu$ and $\sigma \sim \pi/\nu\sqrt{6}$ are the mean and standard deviation of the absorption time. Here $\gamma \approx 0.5772$ is the Euler-Mascheroni constant. Plugging in this transformation and taking $N\rightarrow \infty$, we find
\begin{equation}
\sigma p(\sigma t + \mu) \xrightarrow[]{N\rightarrow \infty} \frac{\pi}{\sqrt{6}} \exp \left(- \gamma - \pi t/\sqrt{6} -e^{-\gamma-\pi t/\sqrt{6}} \right),
\end{equation}
which is precisely the standardized Gumbel distribution.

\begin{figure}[t]
\includegraphics[width=.47\linewidth]{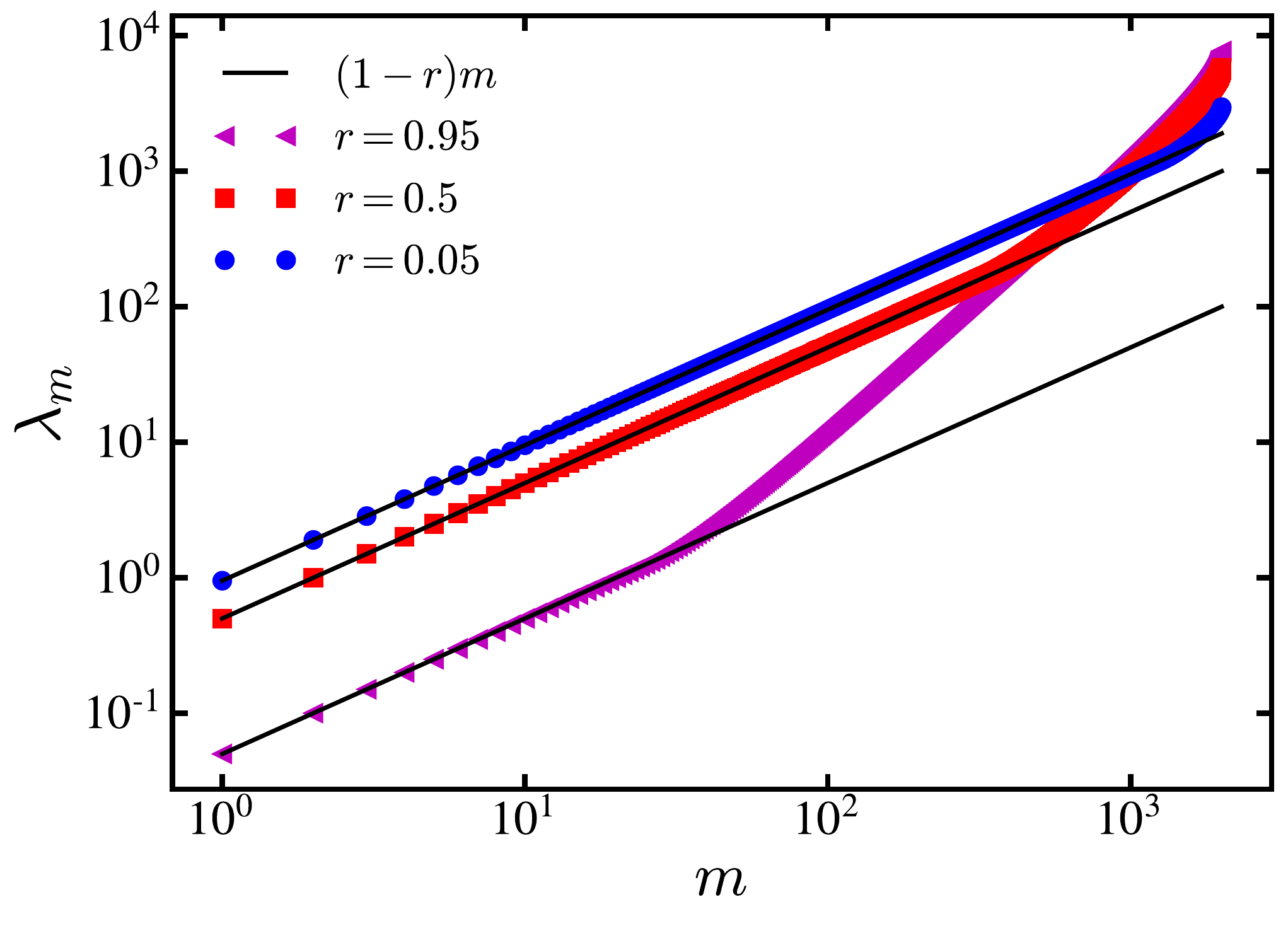} \,
\caption{\label{fig:gumbelEigenvalues}
The eigenvalues of the transition matrix for the canonical model $b_m=rm$, $d_m=m$ with $N=2000$ and $r=0.05$, $0.5$, and $0.95$ plotted on a log-log scale. The black lines show $(1-r)m$ for each value of $r$. The eigenvalues closely follow this linear relation up to a cut-off $m_c$ that is dependent on $r$. Since the leading eigenvalues are linear the absorption-time distribution is Gumbel.
}
\end{figure}

\subsection{Transition matrix spectrum}
The Gumbel distribution also arises if the transition matrix eigenvalues decay linearly. For instance, suppose $\lambda_m = b m$. Then, using Eq.~(\ref{eigenCumulants}) and taking $N\rightarrow\infty$, we have that \begin{equation}
    \tilde \kappa_n = \left((n-1)!\sum_{m=1}^\infty (b m)^{-n} \right) \Bigg /\left(\sum_{m=1}^\infty (b m)^{-2} \right)^{n/2} = (n-1)! \frac{ \zeta(n)}{\zeta(2)^{n/2}},
\end{equation}
which are precisely the cumulants for a standardized Gumbel distribution. The result is unchanged if the dominant eigenvalues are approximately linear, i.e. $\lambda_m \approx b m$ for $m<\alpha N$ where $\alpha$ is a constant $0<\alpha<1$. In this case, the standardized cumulants are still $\tilde \kappa_n = (n-1)!  \zeta(n)/\zeta(2)^{n/2}$ with the larger eigenvalues contributing $\mathcal{O}(1/N)$ corrections that vanish asymptotically.

This second case appears to be what happens in practice: for $N\gg1$ the eigenvalues become linear up to a cutoff. We have carried out numerical calculations of the spectrum for the canonical model from the previous section,  $b_m = r m$, $d_m = m$ for a few values of $r<1$. As shown in Figure~\ref{fig:gumbelEigenvalues}, the leading eigenvalues in the spectrum become equally spaced: $\lambda_m \approx (1-r) m$ for indices below a cutoff $m_c$. Numerical tests indicate $m_c$ is approximately a constant proportion of $N$, i.e. $m_c \approx \alpha(r) N$, where $0<\alpha(r)<1$. Above this cutoff the eigenvalues grow super-linearly. The above calculation illustrates how the Gumbel absorption-time distribution arises in this model from the perspective of the eigenvalue spectrum.

{\color{black} \section{Asymptotic Analysis for the power-law processes}\label{skewedFamily}
In this section we generalize the Gumbel criteria discussed above. Consider Markov processes with transition rates that satisfy $b_m+d_m = f(N) m^p [1+ \mathcal{O}(m/N)]$. Moreover, suppose that $b_{\alpha N} + d_{\alpha N}$ is of order at least $\mathcal{O}(N^p f(N))$ for any $0<\alpha<1$, and $r_m = r + \mathcal{O}(m/N)$ for large $N$. In other words, this process has transition rates that vanish as a power-law $m^p$ near the boundary. In the main text we claimed that $p\leq 1/2$ gives rise to Gaussian absorption times, while $p>1/2$ leads to a skewed family of distributions (whose shape depends on the parameters $p$ and $r$). We rigorously justify these claims in the following subsections.

\subsection{Skewed family for $p>1/2$}
When $p>1/2$ the transition rates decay quickly enough that the process is dominated by slowdown near the boundary (similar to coupon collection), giving rise to skewed distributions. To analyze this case, we can apply similar asymptotic analysis to that given in Section~\hyperref[gumbelBounds]{S4.A} for the Gumbel class. Repeating the bounds in Eqs.~(\ref{bound1}) and (\ref{bound2}) for the power-law process, we find that 
\begin{align}\label{boundsMP}
\sum_{\substack{1\leq i_1 \leq i_2 \leq \dots \leq i_{n-1} \mathstrut \\ \mathstrut \alpha N \leq i_{n-1} \leq i_n \leq N}} \frac{w_{i_1i_2 \cdots i_n}^n(\{r_j\})}{(b_{i_1}+d_{i_1})\cdots(b_{i_n}+d_{i_n})}  &= \mathcal{O}(f(N)^{-n} N^{1-2p})\\
\sum_{\substack{i_1 \leq i_2 \leq \dots \leq i_{n} \mathstrut \\ \mathstrut 1 \leq i_{1} \leq \alpha N \leq i_n \leq N}} \frac{w_{i_1i_2 \cdots i_n}^n(\{r_j\})}{(b_{i_1}+d_{i_1})\cdots(b_{i_n}+d_{i_n})} &= \mathcal{O}(f(N)^{-n} N^{-p}).
\end{align}
As long as $p>1/2$, these terms are each asymptotically dominated by the indices near $0$, 
\begin{equation}\label{sumAsymptoticsMP}
  \sum_{1\leq i_1 \leq i_2 \leq \dots \leq i_n \leq \alpha N} \frac{w_{i_1i_2 \cdots i_n}^n(\{r_j\})}{(b_{i_1}+d_{i_1})\cdots(b_{i_n}+d_{i_n})} \sim  \frac{1}{f(N)^n}  \sum_{1\leq i_1 \leq i_2 \leq \dots \leq i_n \leq \alpha N} \frac{w_{i_1i_2 \cdots i_n}^n(r)}{i_1^p i_2^p \cdots i_n^p},
 \end{equation}
 which are at least of order $\mathcal{O}(f(N)^{-n})$. Similar to the Gumbel class, the absorption process is dominated by the slow behavior near the absorbing state, where the transition rates decay. Extending the upper limit on the sum from $\alpha N$ to $N$ (which only adds subdominant terms), we find that the cumulants $\kappa_n$ satisfy 
 \begin{equation}\label{mpApprox}
\kappa_n(\{r_j\}, N) \sim \frac{1}{f(N)^n}  \sum_{1\leq i_1 \leq i_2 \leq \dots \leq i_n \leq N} \frac{w_{i_1i_2 \cdots i_n}^n(r)}{i_1^p i_2^p \cdots i_n^p}.
\end{equation}
Notice that this asymptotic formula for the cumulants is identical to Eq.~(4) in the main text, but with the denominator $(i_1 i_2 \cdots i_m)$ raised to the power $p$. Thus, we have shown the absorption-time cumulants for a general Markov process, with $b_m+d_m = f(N) m^p [1+ \mathcal{O}(m/N)]$, are asymptotic to those for the minimal model $b_m = r d_m = r m^p$ (after rescaling time so that $f(N) = r+1$). The absorption-time distributions for the minimal model were explored numerically in Figs.~1(d) and 2 in the main text. For $p>1/2$, we find a family of distributions that become more skewed as $p$ increases. The shape of the distributions depends subtly on $r$ except when $p=1$, where the distribution is Gumbel, as revealed by our analysis above.

\subsection{Gaussian distributions for $p\leq 1/2$}\label{powerLawGaussian}
To show the normality of the absorption-time distribution for $p\leq 1/2$, we show that the variance $\kappa_2$ diverges at least as fast as the higher-order cumulants. Using the asymptotic estimate from the previous section, we have that
\begin{equation}\label{mpGaussianGeneral}
\kappa_n(\{r_j\}, N) \sim \frac{1}{f(N)^n}  \sum_{1\leq i_1 \leq i_2 \leq \dots \leq i_n \leq N} \frac{w_{i_1i_2 \cdots i_n}^n(r)}{i_1^p i_2^p \cdots i_n^p} + \mathcal{O}(f(N)^{-n} N^{1-2p}).
\end{equation}
As noted above, when $p<1/2$, we can not guarantee that the first term is dominant. If the second term is dominant than the cumulants scale like $\kappa_n\sim c_n f(N)^{-n} N^{1-2p}$. Then the standardized cumulants asymptotically vanish, $\tilde \kappa_n = \kappa_n/\kappa_2^{n/2} \propto N^{(2-n)(1-2p)/2} \to 0$ as $N\to\infty$. Hence, the absorption-time distribution is asymptotically normal. On the other hand, if the first term in Eq.~(\ref{mpGaussianGeneral}) is dominant, we can show the distribution is still Gaussian. Since the weight factors fall off exponentially away from the diagonal, the diagonal terms are asymptotically dominant. Using the fact that $w_{i_1i_2 \cdots i_n}^n(\{r_j\}) = (n-1)! V_{ii}^n \geq 1$, when $i_1=i_2=\cdots=i_n\equiv i$ together with Eq.~(\ref{visitNumbers}), we have
\begin{equation}\label{mpGaussianSimple}
\kappa_n(\{r_j\}, N) \sim \frac{(1+r)^n}{(1-r)^n f(N)^n}  \sum_{i=1}^N \frac{(1-r^i)^n}{ i^{np}}.
\end{equation}
Notice that when $p=1/2$, the sum in this expression diverges as $N\to\infty$ for the variance ($n=2$), but converges for the higher-order cumulants ($n>2$). More generally, for any $p\leq1/2$, it is straightforward to show that the sum in Eq.~(\ref{mpGaussianSimple}) always diverges faster with $N$ for the variance than for the higher-order cumulants. As above, this scaling leads to $\kappa_n/\kappa_2^{n/2} \to 0$ for large $N$, so that the distribution asymptotically approaches a Gaussian. Figures 1(d) and 2 in the main text show this result is confirmed in numerical simulations: the distribution for $p\leq 1/2$ looks approximately normal and the skew approaches 0.
}

\section{Example models}\label{exampleModels}

{\color{black} 
 In this section we provide details of the evolutionary game, SIS, logistic, and autocatalytic chemical reaction models, each of which exhibit Gaussian, Gumbel, or exponential absorption-time distributions in different parameter regimes. Parameters used for the simulations presented in the main text are provided in Table~\ref{parameterTable}.
}

\subsection{Evolutionary games}

In the main text, we present absorption-time distributions measured via simulations of a two-strategy evolutionary game. In this game, two types of individuals, mutants (M) and wild-types (W), compete and have \emph{frequency dependent} fitness, which means that an individual's fitness depends on the identity of its neighbors. This dependence is encoded by the payoff matrix,
 \begin{equation}\label{payoff}
 \begin{array}{c|cc}
 & M & \,\,\, W \\
   \hline
  
  M&  a & \,\,\,b \\
   W&  c &\,\,\, d
 \end{array}
 \end{equation}
 For example, a mutant (M) with 2 mutant neighbors and 3 wild-type neighbors will have payoff $\pi = 2 a + 3 b$. The fitness is then $\exp(\beta \pi)$, where the parameter $\beta$, the selection intensity, controls how strongly payoff influences fitness. This choice is known as the exponential fitness mapping \cite{ashcroft2016statistical}; we note that other fitness mappings do not change the qualitative behavior discussed below. The dynamics of the model are as follows: an individual is chosen randomly, proportional to their fitness. The selected individual gives birth to an offspring of the same time (M or W) which replaces a random neighbor (selected uniformly). We will let $m$ denote the number of wild-types in the population. Thus, when $m=0$, the mutants have taken over the population (in the jargon, the mutation becomes fixed). We focus on cases in which the mutation becomes fixed, ignoring those when the mutation dies out (which have infinite absorption time). We consider evolution in two types of network populations: a one-dimensional (1D) ring of individuals and a well-mixed (complete graph) population. Each exhibits different absorption-time behavior.

\begin{table*}{\color{black}
\caption{\label{parameterTable} Parameter choices for the simulations used to measure absorption-time distributions shown in Figures~1(a)-(c) and 3(b) in the main text. See Section~\ref{exampleModels} for model and parameter definitions. Evolutionary games use well-mixed population structure except in Figure~1(a). In Figure~1(c) the relative weighting of the convolution of Gumbel distributions is $s=(1 + e^{\beta(c-a)})/(1+ e^{\beta(b-d)}) \approx 0.73$ for both sets of parameters.}
\begin{ruledtabular}
\def\arraystretch{1.2}
\begin{tabular}{ccc}
 
   Figure& Model & Parameters \\  \hline 
 1(a) &1D Evolutionary Game& $\beta=1$, $a=2$, $b=4$, $c=1$, $d=0.1$   \\ \hline
  \multirow{3}{*}{1(b)}& SIS Model& $\Lambda = 0.5$ \\ 
&Logistic Model& $B=0.5$, $K=1$ \\
&Chemical Reaction Model& $k_1 = 1$, $k_2 = 0.75$, $k_3 = 1.25$  \\ \hline
\multirow{2}{*}{1(c)}& Evolutionary Game (black circles) & $\beta=1$, $a=1$, $b=0.5$, $c=0.8$, $d=0.1$ \\ 
&Evolutionary Game (red circles)& $\beta=2$, $a=0.3$, $b=1.3$, $c=0.06$, $d=1.2$ \\ \hline
\multirow{4}{*}{3(b)}& SIS Model& $\Lambda = 1.4$ \\ 
&Logistic Model& $B=1.4$, $K=1$ \\
&Chemical Reaction Model& $k_1 = 1$, $k_2 = 1.35$, $k_3 = 0.14$  \\ 
&Evolutionary Game& $\beta=1$, $a=1$, $b=1.5$, $c=1.2$, $d=1$ \\

\end{tabular}
\end{ruledtabular}}
\end{table*}

\subsubsection{1D ring population structure}

First we consider individuals connected in a 1D periodic ring \cite{altrock2017evolutionary}. Assuming a single initial mutant, the mutant population grows as a connected chain. Any changes in the population must occur at the boundary between mutants and wild-types. The boundary mutants and wild-types have payoff $a+b$ and $c+d$ respectively (they have one of each type as a neighbor). Thus the probability $b_m$ of removing a mutant, and the probability $d_m$ of adding a mutant,  are given by 
\begin{equation}
b_m = e^{\beta (c+d)}/F_m \,, \quad \quad d_m = e^{\beta (a+b)}/F_m \,, \quad \quad \text{for} \,\, 1<n<N-1,
\end{equation}
where $F_m$ is the average fitness:
\begin{equation}
F_m = 2 e^{\beta(a+b)} + (N-m-2) e^{\beta 2 a} + 2 e^{\beta (c+d)} + (m-2) e^{\beta 2 d}.
\end{equation}
The rates are slightly different for $m=1$ and $m=N-1$ \cite{altrock2017evolutionary}. For example, when $m=1$ there is a single wild-type with two mutant neighbors. These transition rates are:
\begin{align}
    b_1 &=\frac{e^{\beta2c}}{2e^{\beta(a+b)}+(N-3) e^{\beta2a} +  e^{\beta2c}}\quad &&\quad \,\, d_1=\frac{e^{\beta(a+b)}}{2e^{\beta(a+b)}+(N-3) e^{\beta2a} +  e^{\beta2c}}\\
    b_{N-1} &= \frac{e^{\beta(c+d)}}{e^{\beta2b}+2 e^{\beta(c+d)} + (N-3) e^{\beta2d}}  \quad &&d_{N-1}= \frac{e^{\beta 2 b}}{e^{\beta2b}+2 e^{\beta(c+d)} + (N-3) e^{\beta2d}}.
\end{align}
For large $N$, however, these changes to the transition rates do not affect the absorption-time distribution. One can check that these transition rates satisfy the requirements of the Gaussian universality class if $(a+b)>(c+d)$.

\subsubsection{Well-mixed population}

If the population is well-mixed, every individual has contact with every other, and hence their fitness depends simply on the fraction of mutants in the population. The payoffs (per contact) for mutants and wild-types respectively are $\pi_M = a (N-m-1)/(N-1) + b m/(N-1)$ and $\pi_W = c (N-m)/(N-1) + d (m-1)/(N-1)$, where again $a$, $b$, $c$, and $d$ are elements of the payoff matrix Eq.~(\ref{payoff}) and $m$ is the number of wild-types in the population. The rates at which the wild-type population increases or decreases are \cite{ashcroft2016statistical}
 \begin{equation}\label{gameCompleteRates}
 b_m = \frac{m \,e^{\beta \pi_W} }{m \, e^{\beta \pi_W} + (N-m) \,e^{\beta \pi_M}} \frac{(N-m)}{N-1}, \quad \quad  d_m = \frac{(N-m) \,e^{\beta \pi_M} }{m \, e^{\beta \pi_W} + (N-m) \,e^{\beta \pi_M}} \frac{m}{N-1}.
 \end{equation}
For the birth (death) rate the first fraction represents the probability of choosing a wild-type (mutant) to give birth, while the second fraction is the probability of the offspring replacing a mutant (wild-type) in the populations. 

Probability flows toward the absorbing state ($r_m<1$) if $b>d$ and $a>c$. From the transition probabilities it is clear $b_m+d_m$ decays linearly near both $m=0$ and $m=N$. Expanding around these points, $b_m + d_m = f(N) (N-m) + \mathcal{O}(m)$ and $b_m + d_m = \tilde{f}(N) (N-m) + \mathcal{O}((N-m)^2)$. Our theory predicts the distribution will be a convolution of Gumbel distributions with relative weighting $s = \lim_{N\rightarrow \infty} f(N)/\tilde{f}(N)$. Taking this limit for the transition probabilities in Eq.~(\ref{gameCompleteRates}) we find
\begin{equation}
s =\frac{1 + e^{\beta(c-a)}}{1+ e^{\beta(b-d)}}.
\end{equation}

\subsection{SIS model}

The stochastic susceptible-infected-susceptible (SIS) model of epidemiology \cite{jacquez1993stochastic} describes the spread of an infectious disease through a population. The population is broken into two groups, those susceptible to the disease and those currently infected. This model describes diseases that do not confer immunity following recovery (or the immunity lasts only for a short time compared to the time scale on which the disease spreads). The rate per contact at which the disease is transmitted between individuals is $\Lambda/N$, and we set the time-scale so that the recovery rate is $1$. Letting $m$ represent the number of infected individuals, there are $m(N-m)$ contacts between infected and susceptible people in a well-mixed population. The $m$ infected individuals each recover at rate $1$. Thus the rates at which the infected population increases and decreases are respectively
 \begin{equation}
 b_m = \Lambda m (1-m/N), \quad \quad d_m = m.
 \end{equation}
 This system has the vanishing transition probabilities near $m=0$, indicating coupon collection behavior (it is also straightforward to explicitly check it satisfies our requirements for the universality class as long as $\Lambda<1$). Our simulations show that it has the expected Gumbel distribution of times for the infection to die out (Fig.~1(b) in the main text).

\subsection{Logistic model}
The stochastic logistic model describes the dynamics and fluctuations of an ecological population \cite{grasman1997local}. The model assumes a constant birth rate $B$ per individual as well as a constant death rate (which we set to $1$ by choosing the appropriate time scale) when the population is sparse. For higher populations, competition between individuals increases the death rate quadratically. The transition rates are 
 \begin{equation}
 b_m = B m, \quad \quad d_m = m + K m^2/N,
 \end{equation}
 where the parameter $K$ controls how strongly competition influences death rates (this parameter is related to the carrying capacity of the ecosystem). Again the transition rates vanish linearly near $m=0$ and this model belongs to the Gumbel universality class as long as $B<1$. 

\subsection{Autocatalytic chemical reaction model}
Our final example model describes a stochastic autocatalytic chemical reaction \cite{doering2007asymptotics},
\begin{equation}
2 X + A \xrightleftharpoons[k_2]{k_1} 3 X, \quad \quad X \xrightharpoonup{k_3} B,
\end{equation} 
where $k_i$ are the reaction rates. The concentrations of species $A$ and $B$ are fixed at saturation levels and we want to describe the dynamics and fluctuations of $m$, the number of particles of species $X$. This is a variation of the Schl\"{o}gl model where the reaction $X\rightarrow B$ is irreversible and the reactions cease when no particles of $X$ remain. Applying our results to this model we will classify the distribution of reaction times: how long does the reaction proceed before the supply of $X$ is exhausted.

The birth-death transition rates for the reaction given above are
\begin{equation}
b_m = \frac{k_1}{N} m(m-1), \quad \quad d_m = \frac{k_2}{N^2} m(m-1)(m-2) + k_3 m.
\end{equation}
Again, the transition rates decay linearly near the absorbing boundary at $m=0$, indicating the Gumbel universality class; it is straightforward to check that the required expansions hold. The conditions that guarantee $r_m = b_m/d_m<1$ are more intricate. In particular, if $k_3>k_2$, then $r_m<1$ as long as $k_1<k_2+k_3$. On the other hand, if $k_3<k_2$ we need $k_2< 2\sqrt{k_2 k_3}$. With either of these conditions satisfied the autocatalytic reaction times will be Gumbel distributed. Note that this model has an infinite state space: the number of $X$ particles can be any positive integer. As discussed in the main text, we expect our results to apply to this class of models as long as the initial condition is large. The simulation shown in Figure~1(b) indicates this expectation is indeed borne out.

\bibliography{SI-references}